\documentclass{cernrep} 
\usepackage{texnames}
\usepackage[T1]{fontenc}
\usepackage[bookmarks, colorlinks=true, linktoc=page, pdftex, linkcolor=black, citecolor=black, urlcolor=blue]{hyperref}

\usepackage{graphbox}

\usepackage{fancyhdr}
\fancyhfoffset{4 mm}
\fancypagestyle{ARTTITLE}{%
\fancyhf{} 
\chead{\small Proceedings of the 2017 CERN--Accelerator--School course on \it Vacuum for Particle Accelerators, Glumsl\"ov, (Sweden)} 
\lfoot{Available online at \url{https://cas.web.cern.ch/previous-schools}}
\rfoot{\thepage\hspace*{3mm}}
 
}

\begin{document}
\title{Materials \& Properties: Thermal \& Electrical Characteristics}
 
\author {Sergio Calatroni}

\institute{CERN, Geneva, Switzerland}

\begin{abstract}
This lecture gives an introduction to the basic physics of the electrical conductivity of metals, its temperature dependence and its limiting factors. We will then introduce the concept of surface resistance, of high relevance in accelerators for its link with beam impedance and for RF applications, including notions related to the anomalous skin effect. The surface resistance will help establishing a link to heat exchanges between bodies by radiation, and to the concept of emissivity. Thermal conductivity will then be introduced, discussing both its electron and phonon exchange components, and the relevant limiting factors.
\end{abstract}


\maketitle 
\thispagestyle{ARTTITLE}

\section{Introduction}
This lecture introduces some basic concepts of solid state physics and material science which are useful for the design of accelerator vacuum components. Most of the material and the basic concepts are discussed in much more detail in classical solid state physics textbooks  \cite{bib:Kittel}, \cite{bib:A&M}, to which the reader is referred for a more accurate and in-depth description of of all concepts developed in the following.

\section{The electrical conductivity}
The vacuum chamber is the macroscopic solid physical boundary with which particle beams interact, essentially through electromagnetic fields. The electromagnetic response of the vacuum chamber material is thus of prime interest for vacuum system design, since this may have for example an effect on beam stability through the beam coupling impedance, explained elsewhere in these lectures.

The electrical conductivity is the quantity which in itself contains all material properties needed to describe the interaction of the electromagnetic fields produced by the beam with the chamber metallic material. The electrical conductivity is typically indicated with the symbol $\sigma$ and its significance stems directly from Maxwell's equations and the Ohm's law, linking the local current density $J$ to the local electric field $E$ in the material: $J=\sigma E $. In material science and engineering, the electrical resistivity $\rho=1/\sigma$ is also widely employed. The electrical resistivity bears a closer link to the electrical resistance $R$ which is a macroscopic property commonly measured with simple devices. It is useful to recall here that for a simple conductor of uniform cross section the electrical resistance $R$ and resistivity $\rho$ are related by the following equation:
\begin{equation}
	R=\frac{\rho \times length}{cross\ section}
	\label{eq:resistance}
\end{equation}
The electrical resistivity is measured in $\Omega$m, while the electrical conductivity is measured in S/m. The electrical conductivity depends on intrinsic  physical characteristics of the metal:
\begin{equation}
	\sigma=\frac{1}{\rho}=\frac{n e^2 \ell}{m_e v_F}=\frac{n e^2 \tau}{m_e}
	\label{eq:conductivity}
\end{equation}
where $n$ is the conduction electron density, $e$ is the electron charge, $\ell$ is the electron mean free path, $m_e$ is the electron (effective) mass, $v_F$ is the Fermi velocity of the conduction electrons in the metal and $\tau$ is the relaxation time between scattering events of the conduction electrons. We shall underline here the obvious important relation between the electron mean free path, the relaxation time and the Fermi velocity $\ell=\tau v_F $. We will not discuss here how the electrical conductivity is derived from the fundamental properties of a metal, nor we will discuss the details of the electron band structure  model compared to the earlier and perhaps more intuitive empirical  free electron metal model.
We will instead concentrate on how the electron scattering, described either through the mean free path $\ell$ or the relaxation time $\tau$ has a direct impact on the conduction properties and how this is related to macroscopic physical, engineering and metallurgical features and fabrication characteristics of a metal.

The scattering of conduction electrons is better seen in the classical free electron model picture. At equilibrium, the electrons are free to move in the metal with a velocity equal to the Fermi velocity, however in their path their undergo collisions which modify at random their trajectories. These scattering events are due to crystal lattice imperfections (vacancies, dislocations), impurities, grain boundaries, the metal surfaces or can be due to collisions with the phonons. Phonons are the quanta of the lattice thermal vibrations, and phonon-electron collisions have an intrinsic temperature dependence, while all the other processes mentioned are essentially independent of temperature. In first approximation, all these scattering processes obey the so-called Mathiessen's rule, which is usually expressed as:
\begin{equation}
	\rho_{total}(T)=\rho_{phonons}(T)+\rho_{impurities}+\rho_{grain\ boundaries}+\ldots .
	\label{eq:Mathiessen}
\end{equation}
This means that each family of scattering processes can be considered individually, and all contribute additively to the total resistivity. Mathiessen's rule can equivalently be written in the form:
\begin{equation}
	\ell_{total}(T)=(\ell_{phonons}^{-1}+\ell_{impurities}^{-1}+\ell_{grain\ boundaries}^{-1}+\ldots)^{-1}
	\label{eq:Mathiessen.mfp}
\end{equation}

In the following sections we discuss more in details the different scattering processes.

\subsection{Temperature-dependent electron-phonon scattering}
\label{sec:e-ph}
As mentioned, phonons are the quanta of the crystal lattice vibrations. Lattice vibrations have discrete frequencies $\omega$ called normal modes, and in a finite crystal there are $3N$ normal modes, where $N$ is the number of atoms (the factor three being related to the three spatial direction). The temperature of the crystal defines (in a statistical way) how many normal modes are excited. For each normal mode the quantized energy of the excitations is $\hbar \omega$, the phonon energy, and many phonons can be excited for each normal mode. (This is equivalent to saying that the oscillations of the atoms in the crystal lattice have amplitudes which increase with increasing temperature).
The maximum angular frequency of the normal modes $\omega_D$ is linked to the temperature $\Theta_D$ by the relation $k_B\Theta_D=\hbar \omega_D$ and these are called the Debye frequency and temperature respectively. The Debye temperature has the following expression:
\begin{equation}
	\Theta_D=\frac{\hbar v_s}{k_B}\left(6\pi^2\frac{N}{V}\right)^{1/3}
	\label{eq:Debye}
\end{equation}
where $v_s$ is the speed of sound and  $V$ the crystal volume. In practice, above the Debye temperature all normal modes are excited, and phonons of all possible frequencies are present in the crystal.

The temperature-dependent component of the electrical resistivity depends on the electron-phonon scattering process, and can be described by the Bloch-Grüneisen equation as a function of the Debye temperature:
\begin{equation}
	\rho_{phonons}(T)=K \left(\frac{T}{\Theta_D}\right)^{-5} \int_{0}^{\frac{T}{\Theta_D}} \frac{x^5}{(e^x-1)(1-e^{-x})}dx=K\left(\frac{T}{\Theta_D}\right)^{-5}J_5\left(\frac{T}{\Theta_D}\right)
	\label{eq:B-G}
\end{equation}
where $K$ is a constant for a given metal, and tabulated values for the function $J_5(T/\Theta_D)$ can be easily found (the index 5 is a convention indicating that for some classes of materials other functional dependencies may be possible).

\begin{figure}
	\centering\includegraphics[width=.5\linewidth]{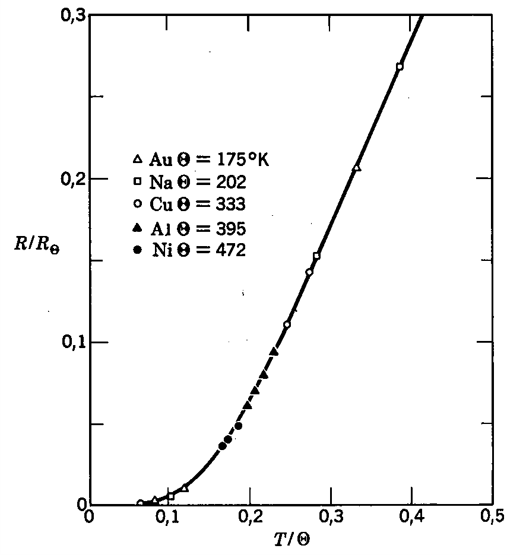}
	\caption{Theoretical temperature change of the resistivity relative to its value at the Debye temperature, compared to the experimental values of a few metals. Debye temperatures of the selected metals are reported in the inset.}
	\label{fig:B-G}
\end{figure}

In \Fref{fig:B-G} we can see the theoretical change of resistivity as a function of temperature (relative to the Debye temperature) compared to the experimental values of a few metals, showing clearly a unified behaviour.
At temperatures $T \gg \Theta_D$ the phonon resistivity is proportional to $T$, while at temperatures $T \ll \Theta_D$ the phonon resistivity decreases proportionally to $T^5$.

In practice, the electron mean free path $\ell$ (or equivalently the relaxation time $\tau$) \textit{increases} with \textit{decreasing} temperature.

\subsection{Other sources of electron scattering}
It is found experimentally that the other main components of the electrical resistivity do not depend on temperature. Among these we can identify a few general categories: impurities, crystal defects (dislocations, vacancies), internal or external surfaces (\ie grain boundaries or the physical surfaces of a specimen).

\begin{figure}
	\centering\includegraphics[width=.5\linewidth]{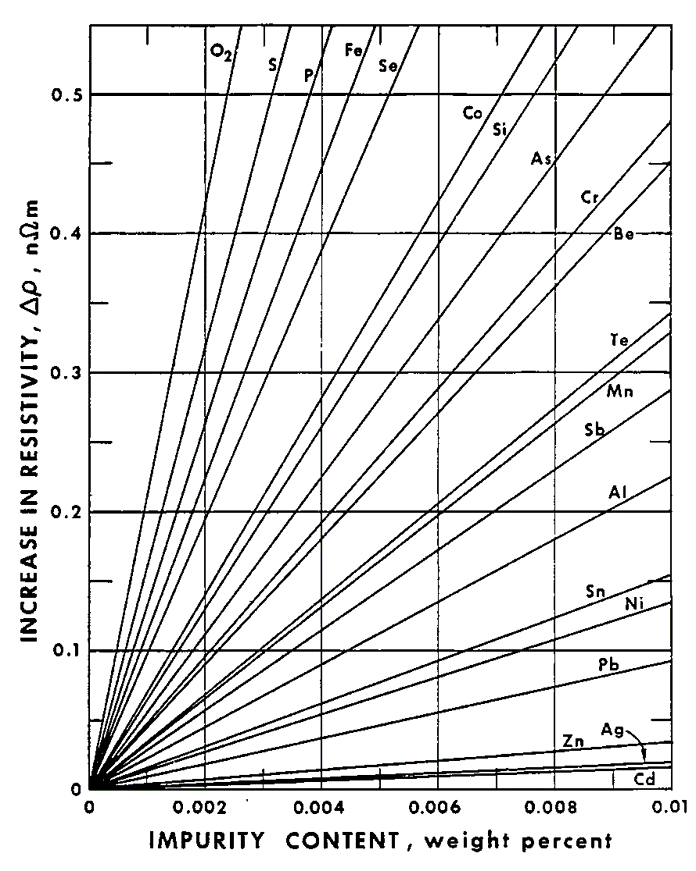}
	\caption{Increase of the electrical resistivity of copper as a function of the concentration of different impurities. For comparison, the electrical resistivity of copper at room temperature is about 15.5~n$\Omega$m}
	\label{fig:Cu_imp}
\end{figure}

\begin{figure}
	\centering\includegraphics[width=.5\linewidth]{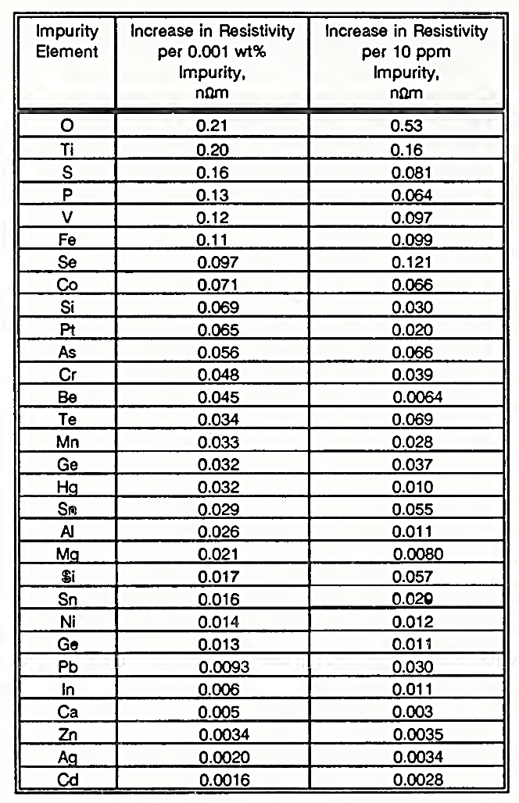}
	\caption{Increase of the electrical resistivity of copper as a function of the concentration of different impurities. It is assumed a concentration of $10^{-5}$ either in weight (first column) or atomic (second column).}
	\label{fig:Table}
\end{figure}

Impurities, up to the concentration of a few atomic \%, have an effect on resistivity which is directly proportional to their concentration. This effect is illustrated as an example in the case of copper in \Fref{fig:Cu_imp}, with the numerical values summarized in the table of \Fref{fig:Table}.

\begin{figure}
	\centering\includegraphics[width=.7\linewidth]{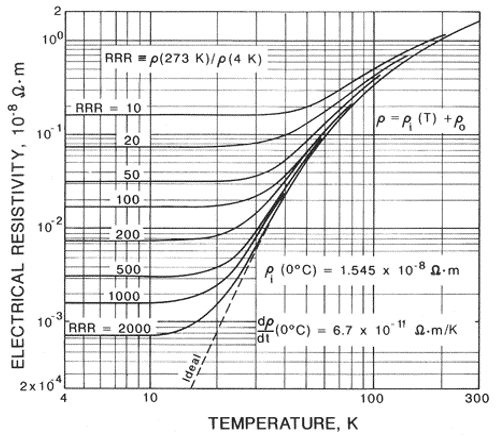}
	\caption{Temperature dependence of the electrical resistivity of copper for different values of the RRR}
	\label{fig:Cu_RRR}
\end{figure}

It is important to appreciate the relative importance of the phonon and of the impurity contributions to the electrical resistivity, and we use the special case of copper for illustrating this. The electrical resistivity of copper for different impurity levels is illustrated in \Fref{fig:Cu_RRR}. The electrical resistivity at room temperature ($\approx 300$~K) is of the order of 15.5~n$\Omega$m, and from the Bloch-Grüneisen model we can see that the temperature dependent part becomes negligible at low temperatures due to the term ($\Theta_D = 333$~K for copper). From the table of \Fref{fig:Table} we can see that a concentration of 10~ppm (atomic) of oxygen would give rise to a resistivity of 0.53~n$\Omega$m, a value about 30 times lower than the approximate value of room temperature resistivity mentioned before.

In fact the ratio of resistivities between room and cryogenic temperature (typically at the boiling liquid helium temperature of 4.2~K) is often used as a global measurement of purity for a metallic specimen, and it is called Residual Resistivity Ratio (RRR). In detail, RRR is defined as:
\begin{equation}
	RRR=\frac{\rho_{total}(300 K)}{\rho_{total}(4.2 K)}\approx \frac{\rho_{phonons}(300 K)}{\rho_0}
	\label{eq:RRR}
\end{equation}
where $\rho_0$ includes all temperature-independent contributions, and with the assumptions that $\rho_{phonons}(300 K) \gg \rho_0$ and that $\rho_{phonons}(4.2 K) \ll \rho_0$. With the prior knowledge that the low-temperature resistivity is limited only by impurities, the RRR could be used for a direct measurement of impurity content, making use of data such as those of \Fref{fig:Table} and of $\rho_{phonons}(300 K)$ (assuming in fact $\rho_{phonons}(300 K)\approx \rho_{total}(300 K)$). In addition, recalling that for a simple conductor of uniform cross section the electrical resistance $R$ and resistivity $\rho$ are related by \Eref{eq:resistance}, we can deduce that 
\begin{equation}
	RRR=\frac{\rho_{total}(300 K)}{\rho_{total}(4.2 K)}=\frac{R_{total}(300 K)}{R_{total}(4.2 K)}
	\label{eq:R_vs_rho}
\end{equation}
and thus the measurement of RRR can be performed by just measuring the electrical resistances, an accurate measurement of the geometry not being necessary. In fact, it can be demonstrated that the RRR of objects of any shape can be measured by just measuring the ratio of the electrical resistances at room and cryogenic temperatures.

\begin{figure}
	\centering\includegraphics[width=.4\linewidth]{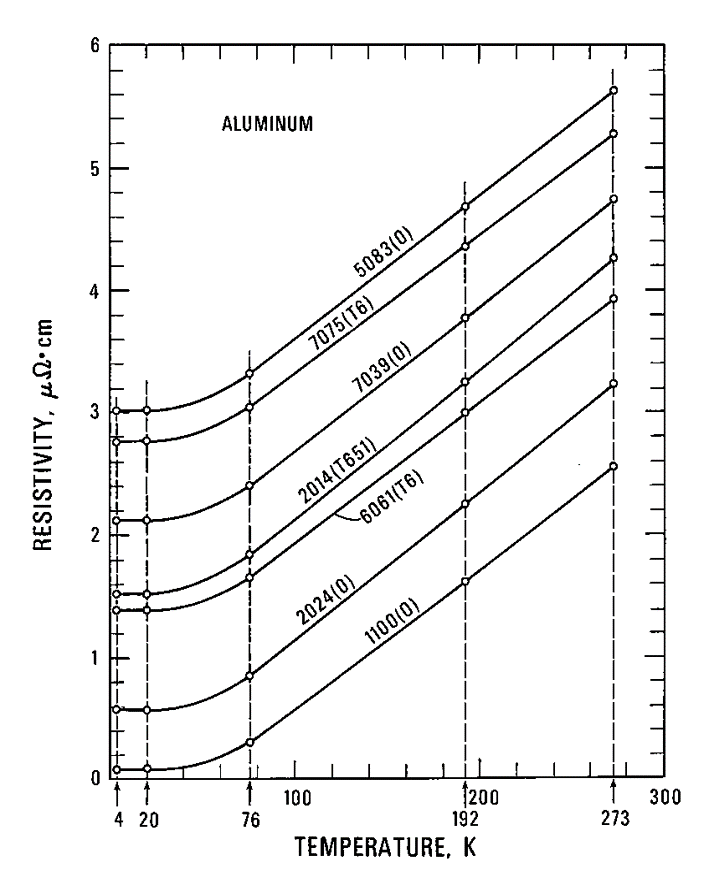}\includegraphics[width=.4\linewidth]{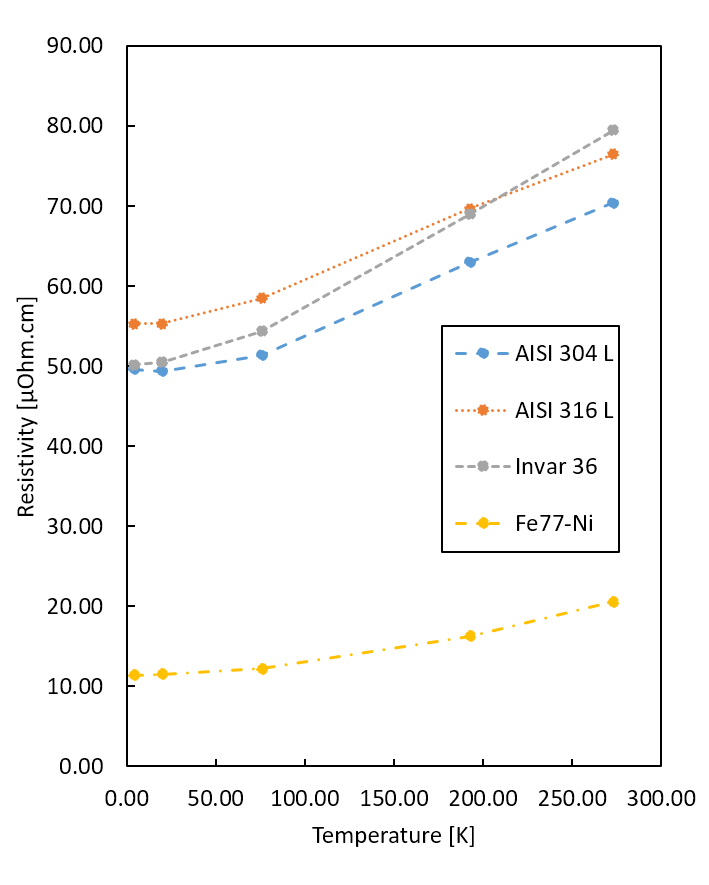}
	\caption{Temperature dependence of the electrical resistivity of some selected aluminium (left) and stainless steel (right) alloys}
	\label{fig:Alloys}
\end{figure}

Alloys can intuitively be considered a special case, where the concentration of one "impurity" may reach levels up to 50\% as for example in binary alloys. The amount of "impurities" is then so large that their effect dominates the electrical resistivity, overshadowing the temperature-dependent part. (In such a case the lattice, the number of conduction electrons and the phonon spectra might also be heavily modified from those of the base metal, but we will not discuss these aspects.) The electrical resistivity of alloys shows then very little changes with temperature, often with RRR values barely larger than one. This is illustrated for some stainless steel and aluminium alloys in \Fref{fig:Alloys}. Special cases are however some metallic crystalline compounds, which may have high values of RRR due to their ordered lattice. This is typically the case of some superconducting compounds used in accelerator technology, such as $\text{Nb}_3\text{Sn}$.

It is important to have an understanding of the order of magnitude of the electron mean free path. Taking again copper as an example, we can first calculate the electron relaxation time from the electrical resistivity at room temperature (see \Eref{eq:conductivity}), using tabulated values for the material parameters and assuming one conduction electron per copper atom (this simple exercise is left to the reader). The resulting value for the electron relaxation time at room temperature is $\tau\approx2.5\times10^{-14}$~s, and knowing that the Fermi velocity is approximately $v_F\approx1.6\times10^6$~m/s, we can estimate that the electron mean free path at room temperature is about $\ell\approx4\times10^{-8}$~m. Most metals have mean free path values in this range at room temperature. We should also consider that it is rather commonplace to manufacture accelerator components out of high-conductivity copper (so-called Cu-OFE standardized as UNS Cu10100), which can have RRR$\approx$100. In this case the mean free path at cryogenic temperature would be of about 4~$\mu$m.

A mean free path value of a few micrometres can be of magnitude comparable to the grain size of typical metallic specimens. Thus we cannot make the simple assumption that the electrical resistivity at cryogenic temperature is dominated by impurities, but we must take into account also the scattering effect of the grain boundaries (as well as also the effect of all other crystallographic features). Typical metallurgical preparations (cold working, annealing) as well as all manufacturing processes (machining, welding, \etc) have certainly an effect on the grain size, and on the amount of dislocations, vacancies, \etc. It is not the purpose of this lecture to go in the details of this, but as an example we can intuitively mention that in the calculation of the total mean free path of Eq.~\ref{eq:Mathiessen.mfp} the contribution of the grain size to the total mean free path could in fact be the grain size itself, thus $\ell_{grain\ boundaries}=grain\ size$.

In fact also external surfaces have a similar effect. In other terms, for very thin films, the top and bottom surfaces can become the limiting factor of the mean free path (in particular at low temperature, with a vanishing phonon contribution). It is worth mentioning that, in this regime, the simultaneous knowledge of thickness and electrical resistivity allows calculating the product $\displaystyle\rho\ell=\frac{m_e v_F}{n e^2}$ (where $\ell$ in this case is just the film thickness), which depends only intrinsic material parameters, and from which the Fermi velocity could be extracted.

\subsection{Magnetoresistance}
\begin{figure}
	\centering\includegraphics[width=.5\linewidth]{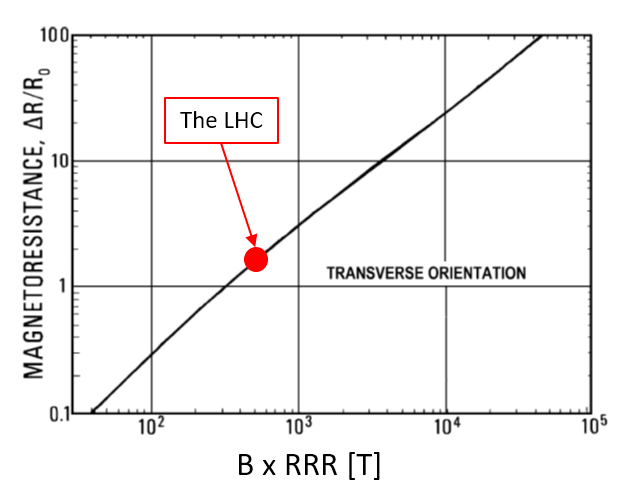}
	\caption{Magnetoresistance of copper illustrated as a "Kohler's plot". The operating conditions inside the superconducting dipoles of the LHC (relevant for the beam screen) are indicated by the arrow.}
	\label{fig:Kohler}
\end{figure}

Magnetoresistance happens when a metal is subject to a strong magnetic field, of the order of a few T, as is typically the case inside a superconducting magnet. In such cases, the electrical resistivity is enhanced by the magnetic field, a phenomenon which is relatively more pronounced in pure materials. This is described by the so-called Kohler's law which says that the relative increase of electrical resistivity is:
\begin{equation}
	\frac{\Delta \rho}{\rho_0}=F\left(\frac{B}{\rho_0}\right) \approx F(B \times RRR)
	\label{eq:Kohler}
\end{equation}
recalling from Eq.~\ref{eq:RRR} that $RRR\approx1/\rho_0$. Kohler's law is illustrated for copper in \Fref{fig:Kohler} that shows the typical linear behaviour measured at high field (the behaviour is quadratic with $B$ at very low magnetic fields, which we will not discuss).

In order to propose a simple explanation of magnetoresistance, we have first to clarify the concept of drift velocity of the electrons. The Fermi velocity $v_F$ is is exceedingly high in metals: in the case of Cu as we mentioned earlier it is of the order of 0.5\% of the speed of light! In order to carry an electric current, the conduction electrons in a metal have obviously to collectively drift, with a speed $v_d$ which can be easily calculated from the current density $J=en_ev_d$. Applying Ohm's law $J=\sigma E$ and assuming the hypothetical reasonable practical case of a copper wire of 1~m of length which carries a current resulting in an applied potential of 1~V, we obtain a drift velocity $v_d\approx4\times10^{-4}$~m/s, orders of magnitude smaller than $v_F$ (the details of this example are left to the reader). The drift velocity, and the resulting travelled distance in the drift direction, are thus negligible compared to the Fermi velocity.

In a magnetic field $B$, electron trajectories are bent with a radius described by the cyclotron radius $\displaystyle r_c=\frac{mv_F}{eB}$. If the relaxation time is long, the electrons will be able to travel a relatively long path along a circular trajectory until the first scattering, having travelled in average a length $\ell$ along this circular path. The length of the path projected in the drift direction can thus be dramatically shortened by this circular movement, and under extreme circumstances the electron might even return close to the starting point (the drift velocity being very small). It is indeed the path length projected in the drift direction that enters the calculation of electrical resistivity.
If the relaxation time is short, the trajectory might only be slightly bent before the next collision. In this case, the projected path length in the drift direction is only slightly changed. The \textit{relative} effect of magnetoresistance is thus stronger when the relaxation time is longer, thus it is in fact relevant from an engineering point of view only at low temperature and for ultra-pure materials. We should however note from Eq.~\ref{eq:Kohler} and within the limits of \Fref{fig:Kohler} that the \textit{absolute} increase of resistivity depends only on the magnetic field and not on the mean free path.

\section{The surface resistance}
\label{sec:Rs}
The Ohm's $J=\sigma E$ law describes the relationship between fields and currents \textit{inside} a metal. The surface resistance, in its simplest meaning, describes the relationship between fields \textit{outside} the surface of a metal and the currents \textit{inside} it. Its main purpose in our context is to evaluate the effect of the electromagnetic fields within the accelerator vacuum on the metallic boundaries of the vacuum system, and vice-versa. The surface resistance $R_s$ is defined as the ratio $R_s=E/H$, where $E$ and $H$ are the electric and magnetic components of the wave at the surface of the metal, and is equal to \cite{bib:Stratton}:
\begin{equation}
	R_s=\sqrt{\frac{\rho \mu_0 \omega}{2}} ,
	\label{eq:Rs}
\end{equation}
where $\omega$ is the angular frequency of the wave. The surface resistance is actually the real part of a complex quantity called the surface impedance $Z_s=R_s+i X_s$ which appears naturally when using the complex notation for describing the electromagnetic fields $E$ and $H$. For a normal metal, it is found that $R_s=X_s$.

The concept of surface resistance is valid at high frequencies (\ie in RF and not in DC), in a regime thus where the fields don't propagate into the metal but decay exponentially within the skin depth $\delta$. For the typical case of planar electromagnetic waves parallel to the surface and penetrating perpendicularly into a metal the skin depth is:
\begin{equation}
	\delta=\sqrt{\frac{2\rho}{\mu_0 \omega}} .
	\label{eq:delta}
\end{equation}
We can then note from \Eref{eq:delta} and \Eref{eq:Rs} that: 
\begin{equation}
	R_s=\frac{\rho}{\delta}
	\label{eq:Rs&delta}
\end{equation}

In practical terms, copper at room temperature and at a frequency of 400~MHz has a surface resistance $R_s$ of about 5~m$\Omega$ and a skin depth $\delta$ of about 3~$\mu$m. The surface resistance \textit{increases} with $\sqrt{\omega}$ while the skin depth \textit{decreases} with $\sqrt{\omega}$.

The importance of the surface resistance is immediately understood when considering that the power dissipated per unit surface by an electromagnetic wave in a metallic conductor is equal to
\begin{equation}
	P=\frac{1}{2} R_s H^2 ,
	\label{eq:power}
\end{equation}
where $P$ is the average power dissipated over a period $\tau=1/\nu=2\pi/\omega$ and $\nu$ is the frequency of the wave.

\begin{figure}
	\centering\includegraphics[width=.8\linewidth]{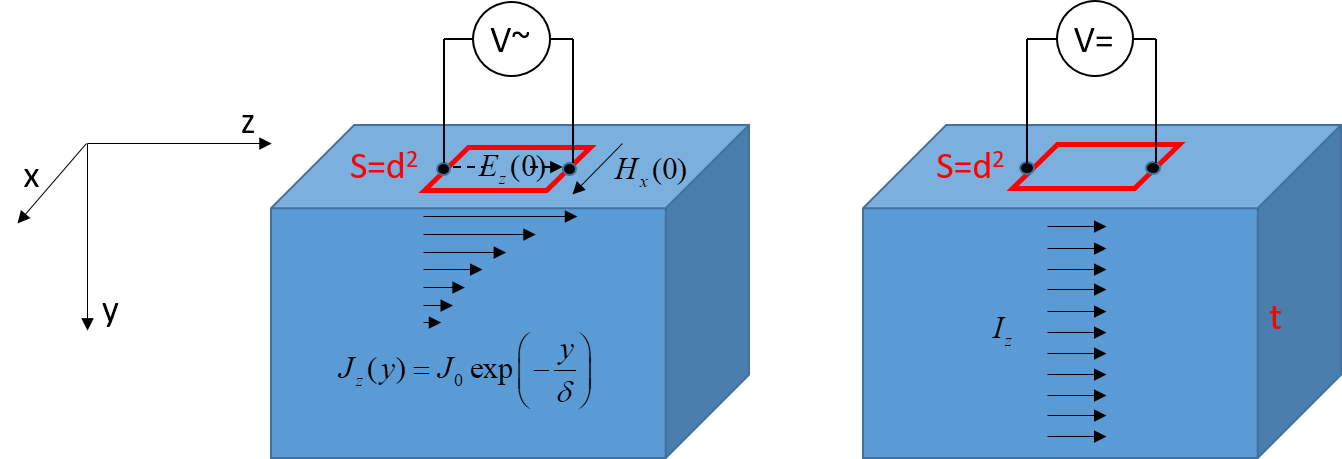}
	\caption{Pictorial illustration of surface resistance $R_s$ in the RF case compared to the square resistance $R_{\square}$ in DC}
	\label{fig:Rsquare}
\end{figure}

It is a useful exercise to compare the surface resistance with the so-called square resistance in DC, see for this purpose \Fref{fig:Rsquare}. The square resistance $R_{\square}$ is defined as the resistance of a thin square slab of side $d$ and thickness $t$, traversed by a current $I$. Recalling \Eref{eq:resistance}, one can easily find that $R_{\square}=\rho/t$, in perfect analogy with $R_s=\rho/\delta$, see \Eref{eq:Rs&delta}. In RF the currents are confined in a surface layer defined by the skin depth $\delta$, thus the resistance is typically higher than in DC. It is left to the reader the exercise of comparing \Eref{eq:power} with its analogous in DC, knowing that in the simple geometry of our example $H=J\times t$, where $J$ is the current density in $\text{A/m}^2$. From this example and Eq.~\ref{eq:power} it should also be clearer the link between \textit{external} fields and \textit{internal} currents established by $R_s$ and mentioned above.

\subsection{Free space and metals}
It is interesting to compare the surface resistance (impedance) with the so-called impedance of free space. This is the ratio $Z_0=E_0/H_0$, where $E_0$ and $H_0$ are the electric and magnetic components of a free wave in vacuum, and is a characteristics of the propagation of waves in free space. Its value is equal to $Z_0=R_0=\sqrt{\mu_o / \epsilon_0}=376.7 \Omega$ (real quantity), and it is immediately apparent the huge difference in magnitude compared to the surface resistance.

We can also recall that in a metal the permittivity can be expressed in complex notation as $\displaystyle \epsilon_{metal}=\epsilon_0 + i \frac{\sigma}{\omega}$. By making the substitution $\epsilon_0 \rightarrow \epsilon_{metal}$ in the expression for $Z_0$ one could derive $Z_s=R_s+i X_s$ with $R_s$ (and $X_s$) equal to \Eref{eq:Rs}. This exercise is left to the more theoretically-minded reader. We  simply underline that the quantity $\sigma / \omega$ is much larger in magnitude than $\epsilon_0$ in normal metals and at RF frequencies (\ie when $\omega \tau \ll 1$ as we will discuss later), and these appearing at the denominator in the expression of $Z_0$ it is clear why the surface impedance is much smaller than its free-space counterpart. 

In more simple terms, the electromagnetic wave induces surface currents in the metal; these currents result in power dissipation ($R_s$) and produce in turn an electromagnetic field which reduces the amplitude of the incident wave; they also change the propagation characteristics of the wave in the metal ($X_s$), which just penetrates over the skin depth, and modify its phase.

\subsection{Surface impedance in particle accelerators}
We have put a strong accent on surface resistance because this quantity is the key link between the material properties of beam-facing components and the beam dynamics.

\subsubsection{Beam impedance}
As discussed extensively elsewhere in these lectures, the concept of beam coupling impedance is of paramount importance in describing particle beam dynamics, and the reader is invited to consult \cite{bib:Wanzenberg} for a detailed description of all the relevant concepts and effects.

\begin{figure}
	\centering\includegraphics[width=.5\linewidth]{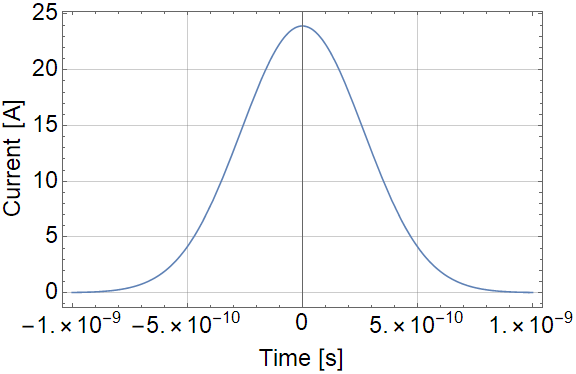}\includegraphics[width=.5\linewidth]{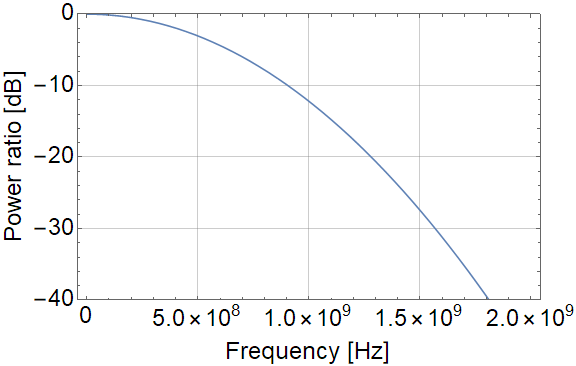}
	\caption{Instantaneous current of LHC-type bunches of $10^{11}$ protons having a Gaussian longitudinal profile with $\sigma=8$~cm (left), and the corresponding \textit{bunch} power spectrum, which is its Fourier transform (right). The full \textit{beam} power spectrum is in fact a collection of line modes under a global envelope having this same shape.}
	\label{fig:PowerSpectrum}
\end{figure}

From the material's perspective, the longitudinal coupling impedance $Z_L$ is a convenient tool for evaluating the power dissipated by the beam induced image currents in the vacuum chamber due to Joule effect. In the idealized case of a circular particle accelerator made of a cylindrical smooth vacuum chamber, it can be expressed in the simplest form as \cite{bib:Sacherer}
\begin{equation} 
	Z_{\text{L}} = \frac{2 \pi R}{2 \pi b} Z_{\text{s}} \: ,
	\label{eq:ZL}
\end{equation}
where $R$ is the radius of the particle accelerator, $b$ is the radius of the vacuum chamber, and $Z_{\text{s}}$ is the surface impedance of the material of the vacuum chamber, as discussed above. Intuitively, \Eref{eq:ZL} is the integration of $Z_{\text{s}}$ over the entire inner surface of the accelerator. The resistive power loss is then equal to
\begin{equation} 
	P_{\text{loss}} = M I_{\text{b}}^2 \operatorname{Re}(Z_{\text{loss}}) \: ,
	\label{eq:Ploss}
\end{equation}
where $M$ is the number of particle bunches in the accelerator, $I_{\text{b}}$ is the bunch current, and $Z_{\text{loss}}$ is the convolution of the longitudinal impedance $Z_{\text{L}}$ with the \textit{beam} power spectrum, whose envelope is illustrated in \Fref{fig:PowerSpectrum} for the case of Gaussian bunches. This is of crucial importance in a cryogenic accelerator such as the LHC and its ongoing upgrade, the High-Luminosity LHC (HL-LHC).

Impedance is of paramount importance for studying beam stability. The accelerator size, its magnet optics, and the particle energy (often determined by a highly relativistic velocity $v \approx c$) define a reference orbit and a rotation frequency for a hypothetical synchronous particle circulating along this reference orbit. In practice the particle bunches, in their motion around the accelerator, perform longitudinal oscillations about the virtual position of the synchronous particle (synchrotron oscillations). They also perform transverse oscillations with respect to the reference orbit, both in the horizontal and vertical planes (betatron oscillations). A large number of factors influence the stability of these oscillations, and the effect of the electromagnetic fields which are excited by the beam in the vacuum chamber (wakefields) is of particular importance. The wakefields act back on the particle beam, and influence both its longitudinal and transverse dynamics, and their effect is of course described in terms of the beam coupling impedance. 
For longitudinal (synchrotron) oscillations this leads to a definition of $Z_{\text{L}}$ identical to \Eref{eq:ZL}. Although longitudinal instabilities can also occur, we discuss here only transverse instabilities. The transverse coupling impedance $Z_{\text{T}}$ can be expressed as \cite{bib:Sacherer}:
\begin{equation} 
	Z_{\text{T}} = \frac{2 R c}{ b^3 \omega} Z_{\text{s}} \: ,
	\label{eq:ZT}
\end{equation}
where $c$ is the speed of light, and $\omega$ is the wakefield frequency. (We should note that $Z_{\text{s}}$ itself has an intrinsic dependence on $\omega$.). The real part of the transverse impedance allows calculating the growth rate of coupled-bunch transverse instabilities, i.e. the exponential growth of the amplitude of the bunch transverse oscillation due to the wakefield created by preceding bunches. The so-called resistive wall transverse instability has a temporal growth rate $\tau$, which exhibits the dependence
\begin{equation} 
	\frac{1}{\tau} \propto \frac{I_{\text{b}} M }{E L} \operatorname{Re}( Z_{\text{T}}^{\text{eff}}) \: ,
	\label{eq:tau}
\end{equation}
where $L$ is the length of the particle bunches and $E$ is the beam energy, and $Z_{\text{T}}^{\text{eff}}$ is calculated for each transverse oscillation mode.
In some cases instability growth times may be equivalent to a few turns only of the particle bunches inside the accelerator.

On the other hand, high-frequency interactions of different oscillation modes within a single bunch, called transverse mode coupling instabilities, set a maximum limit to the bunch current. Provided some other resonant conditions are established, the intensity threshold can also be conveniently described in terms of $Z_{\text{T}}^{\text{eff}}$ as
\begin{equation} 
	I_{\text{b}} \propto \frac{E}{\operatorname{Im}(Z_{\text{T}}^{\text{eff}})} \: .
	\label{eq:intensity}
\end{equation}

It is apparent that in all cases it is most favourable to minimize the surface impedance of the material facing the particle beam.

\subsubsection{RF cavities}
RF cavities are the devices used to accelerate the charged particle beams in an accelerator. They are typically made of copper or of some superconducting material, in order to minimize the energy lost by Joule effect. The reason for this is intimately linked to the surface resistance of the materials used for their manufacturing.

\begin{figure}
	\centering\includegraphics[width=\linewidth]{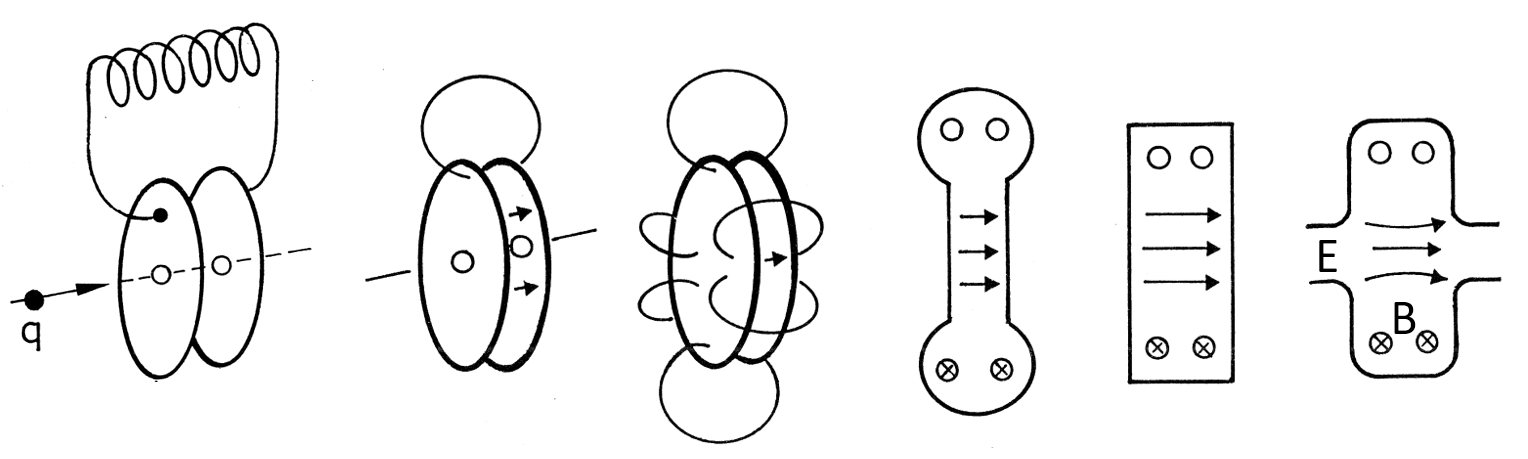}
	\caption{From left to right, metamorphosis of an RLC circuit into a "pillbox" cavity. The electric field $E$ is indicated by arrows, the magnetic field $B$ by crosses.}
	\label{fig:RLC}
\end{figure}

In an accelerating cavity a resonating RF wave is established by feeding RF power from an external generator, and the electric component of the resonating wave is synchronized with the passing particle bunches so to accelerate them. Although most of the RF power should ideally be transferred to the particle beam, in reality a fraction is inevitably lost in the cavity inner walls by Joule effect. This is often characterised by the cavity quality factor $Q_0$. The quality factor is a typical measure of the internal losses of a resonator, be it mechanical, electrical, or an accelerating cavity, and very often cavities are illustrated as a metamorphosis of an RLC circuit as in \Fref{fig:RLC}. The quality factor is defined as:
\begin{equation} 
	Q_0=\frac{f_0}{\Delta f_0}=\frac{2 \pi f_0 U}{P_c} ,
	\label{eq:Q0}
\end{equation}
where $f_0$ is the resonant frequency, $\Delta f_0$ is the width of the resonance, 
 $U$ is the energy stored in the cavity volume and $P_c$ is the power dissipation.

Experimentally, the energy in the cavity decreases as  $U=U_0 \text{exp}(-t/\tau)$ when the RF generator is turned off, and the decay constant is $\tau=1/\Delta f_0$.
The stored energy is $\displaystyle U=\frac{1}{2}\mu_0\int_V |H|^2 dv$ and the power loss, recalling \Eref{eq:power}, is $\displaystyle P_c=\frac{1}{2} R_s \int_S |H|^2 ds$, where the integrals are over the cavity volume and surface respectively. These integrals depend only on the cavity geometry, thus it results:
\begin{equation} 
	Q_0=\frac{\Gamma}{R_s} ,
	\label{eq:Q}
\end{equation}
where the geometry factor $\Gamma$ takes into account all the geometrical details of the cavities, and can be easily evaluated by specialized computer codes. The geometrical factor for a simple cylindrical cavity, whose axis is parallel to the particle beam ("pillbox" cavity, see \Fref{fig:RLC}), can even be evaluated analytically and is $\Gamma$=257~$\Omega$, in the ballpark of typical realistic cavities. Recalling the surface resistance of copper at 400~MHz (typical frequency of accelerator cavities, used for example in the LHC) mentioned above of about 5~m$\Omega$, we can estimate that the quality factor of accelerating cavities is of the order of $Q\approx 50000$. In different terms, the width of the resonance $\Delta f_0$ in this example is about 8 kHz.

\subsection{The anomalous skin effect in normal metals}
\label{sec:ASE}
When cooling down a good conductor such as copper the electrical resistivity decreases, as we have seen earlier. The surface resistance should also decrease as $\sqrt{\rho}$, according to \Eref{eq:Rs}. However, under certain circumstances, the surface resistance ceases to decrease and stabilizes to a plateau. This is not the same phenomenon seen in \Fref{fig:Cu_RRR}: in the latter, the \textit{resistivity} stabilizes at low temperature due to impurities while in the former the \textit{surface resistance} stabilizes, at temperatures where the \textit{resistivity} is still decreasing. This phenomenon is known as the anomalous skin effect, and may have an impact on the performance of some accelerator components at cryogenic temperatures (the LHC beam screen is one such case).

\begin{figure}
	\centering\includegraphics[width=.9\linewidth]{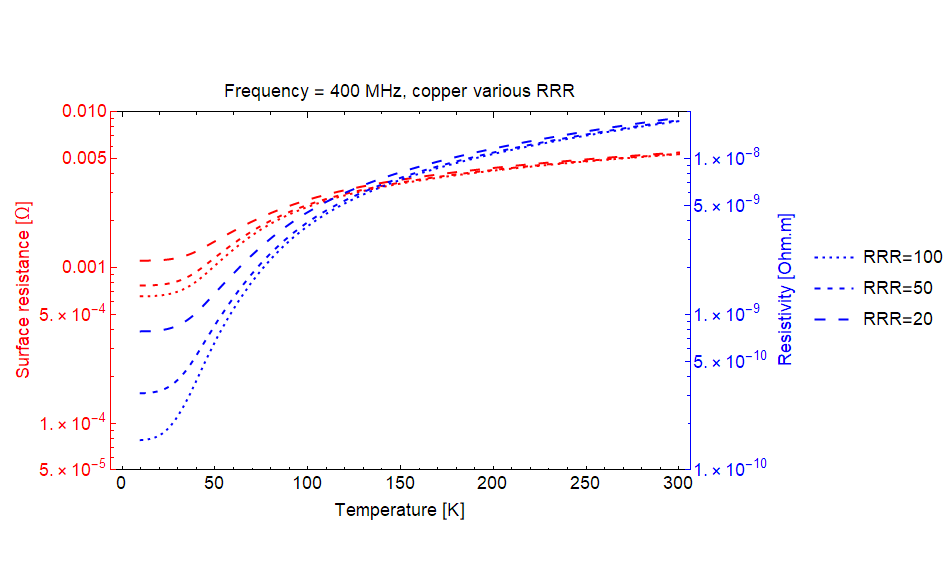}
	\caption{Surface resistance at 400 MHz (left axis and set of curves with the shallower variation) and resistivity (right axis and set of curves with the larger variation) as a function of temperature for copper with different values of RRR}
	\label{fig:Rs_rho_vs_T}
\end{figure}

As discussed in \Sref{sec:e-ph} the mean free path increases when lowering the temperature. At any given frequency $\omega$, the mean free path might eventually become larger than the skin depth, see \Eref{eq:delta}. Remembering that the skin depth is defined by the currents penetrating into the material, and thus shielding its interior from the external fields, we can see that a hypothetical conduction electron which has a trajectory starting from the surface and directed towards the interior of the material might take part to the shielding current in the first part of its path, and then proceed undisturbed into a zero-field region. This does not usually happen, since some electron scattering would always take place within the skin depth, maintaining all electrons participating to the shielding currents within it. In an equivalent way, one could say that in the anomalous skin effect regime only a fraction $\displaystyle n_{eff} \approx n_e \frac{\delta}{\ell}$ of the conduction electrons take part to the shielding currents. It goes beyond the purpose of this lecture to discuss the exact solutions of this problem: the reader might be interested in substituting the value of  $n_{eff}$ for $n_e$ in \Eref{eq:conductivity} and substitute the result in \Eref{eq:delta} to calculate the skin depth $\delta$. It is found that in this \textit{extreme} limit the value of $\delta$ does not depend any more on mean free path and reaches an asymptotic value dependent only on intrinsic material quantities. The surface resistance reaches also an asymptotic value, as discussed earlier, which could be evaluated from \Eref{eq:Rs&delta}. In \Fref{fig:Rs_rho_vs_T} the dependence of the surface resistance and of the resistivity as a function of temperature for copper is reported, where the limited influence of the RRR on the former is clearly visible. We should also mention that the anomalous skin effect may appear at any temperature, for sufficiently high frequency, as obvious inspecting \Eref{eq:delta}.

One could finally note that in the anomalous skin effect regime $R_s \le X_s$, ($X_s=\sqrt{3} R_s$ in the \textit{extreme} limit). We can also underline that even the scattering process of the conduction electrons from the surface of the metal (specular or diffuse) plays a role, in analogy to very thin films.

\subsection{Superconductors}
Superconductors loose completely their electrical resistivity at temperatures below a critical temperature $T_c$. All superconductors of interest for accelerator applications have rather low values of $T_c$, and among these the most used are Nb (9.2~K), NbTi (9.5~K) and $\text{Nb}_3\text{Sn}$ (17.3~K). High-temperature superconductor ceramics such as YBCO and BiSCCO have only limited use, in very specific niches.

In an RF field at $T>0 \ \text{K}$ a superconductor has a small but non-zero surface resistance, theoretically vanishing only at 0~K. This rather surprising phenomenon, linked to the fact that not all conduction electrons are bound in Copper pairs at $T>0 \ \text{K}$, has a very complex theoretical explanation. In practice in the case of niobium, which is the material of choice for the fabrication of superconducting accelerating cavities, either in the form of bulk metal or in the form of thin films coated onto copper structures, the surface resistance is in the range of a few to a few tens of n$\Omega$. This represents five to six orders of magnitude lower than copper at room temperature! Even taking into account the efficiency of the plants needed for refrigeration (for example, 250~W of electrical power for 1~W of cooling power at 4.5~K for the LHC cryogenic plants), the energy saving is extremely advantageous. 

The quality factor $Q_0$ of a superconducting cavity results then to be in the range $10^9 - 10^{10}$, to be compared to a few $10^4$ mentioned before for copper. The bandwidth can then be smaller that 1~Hz, which represents a considerable challenge for the accelerator control systems which have to tune the cavity in phase with the particle beam.  

As a final note, we mention that in superconducting cavities $R_s \ll X_s$, related to the fact that the skin depth is only slightly smaller compared to normal metals and in the order of $\approx 100$~nm, while the surface resistance as mentioned is extremely small.

\section{Heat exchange}
Heat exchanges in vacuum are of course much different compared to heat exchanges in air. A body in vacuum can exchange heat with its surrounding environment only through radiation or through direct contact, since there is no surrounding media enabling conductive or convective heat exchange. Components in vacuum must then be carefully engineered, and the materials adequately chosen, in order for them to properly operate. As a few examples, we can mention beam intercepting device, beam monitoring devices, shielding components, ferrite absorbers, to name just a few which might get heated by interaction with the beam and thus need proper cooling. Cryogenic components, such as vacuum chambers, magnets and RF cavities, require also proper insulation from ambient temperature, the most effective being vacuum insulation. In such a case, heat exchange can be the limiting factor for a proper and energy efficient cooling.

\subsection{Heat exchange by radiation}
\begin{figure}
	\centering\includegraphics[width=.3\linewidth, align=c]{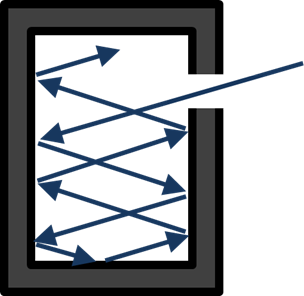}\includegraphics[width=.5\linewidth, align=c]{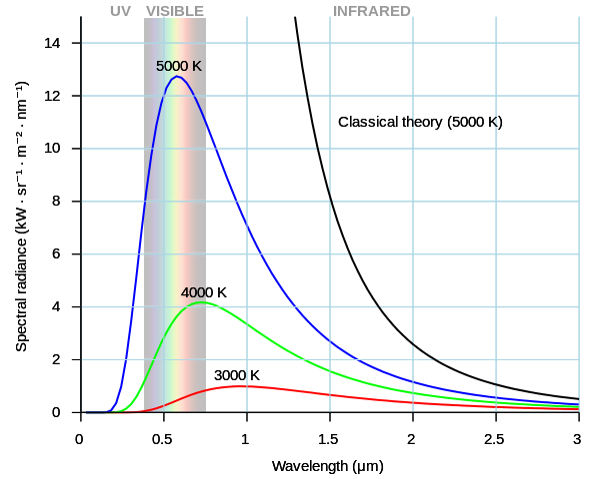}
	\caption{Pictorial illustration of a blackbody (left). Plot of the spectral radiance of a blackbody for different temperatures (right)}
	\label{fig:Blackbody}
\end{figure}

When we speak of radiation in the context of heat exchanges we imply electromagnetic radiation, of wavelengths which may go from low-frequency RF to visible light and beyond. A description of heat exchange by radiation must start by mentioning the so-called blackbody, which is an idealized object, perfectly emitting and absorbing electromagnetic radiation (often schematized as a cavity with a tiny hole), see \Fref{fig:Blackbody}. To explain experiments on blackbody radiation Planck laid the foundations of quantum theory, postulating for the first time that electromagnetic radiation is emitted and absorbed only in discrete quanta (photons). The common experience is that a very hot body can emit light, its colour changing progressively from red to more blueish shades as the temperature increase. This is apparent in \Fref{fig:Blackbody} where the power radiated by an ideal blackbody as a function of wavelength is illustrated for different temperatures, as calculated from Planck's theory. One should note that the peak of the distribution is located at $\approx 3000 \ \mu \text{m} \times \text{K}$, thus the higher the temperature the shorter the wavelength of the peak. At room temperature ($\approx \ 300$~K) the peak is then at $\approx \ 10 \mu$m, in the near infrared. The visible spectrum goes roughly from 400~nm to 750~nm which correspond to a peak of about 7500~K and 4000~K respectively. This is very well described by Planck's theory, and the fact that some light is usually visible starting already from about 1000~K is due to the tail of the Planck distribution. 

As a whole, a blackbody obeys also the Stefan-Boltzmann law for the integral radiated power $P$ per unit area $A$ (power density):
\begin{equation} 
	\frac{P}{A}=s T^4
	\label{eq:StefanBoltzmann}
\end{equation}
with the Stefan-Boltzmann constant $s=5.67\times10^{-8} \ \text{Wm}^{-2}\text{K}^4$.

\begin{figure}
	\centering\includegraphics[width=.4\linewidth, align=c]{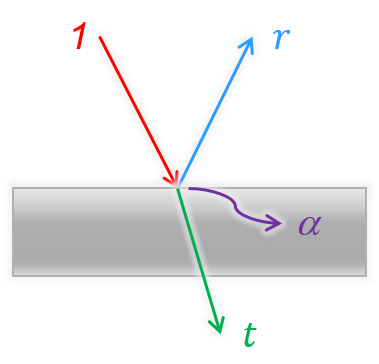}\includegraphics[width=.5\linewidth, align=c]{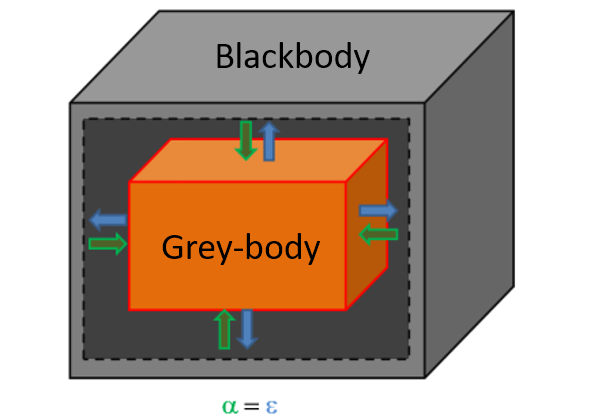}
	\caption{Pictorial illustration of the sum of the different components $r+\alpha+t=1$ (right). Pictorial illustration of a grey body in thermal equilibrium within a blackbody of a given temperature, in which situation $\alpha=\epsilon$, with $t$ being usually negligible (right).}
	\label{fig:Kirchoff}
\end{figure}

A real object radiates power \textit{always} less efficiently than a blackbody for a given temperature, or equivalently for the same radiated power its temperature must be larger than for and ideal blackbody. This is usually expressed through the concept of emissivity:
\begin{equation} 
	\frac{P}{A}=\epsilon s T^4
	\label{eq:StefanBoltzmann2}
\end{equation}
where the emissivity $\epsilon$ is less than unity, and may depend on the radiated wavelength $\lambda$ and on the temperature itself. A real object is often called "grey", in contrast with the ideal blackbody. When a grey body is irradiated, part of the radiation is reflected, part is absorbed, and some may be directly transmitted across (this latter part is usually negligible) \Fref{fig:Kirchoff}. Assuming the incident radiation is equal to unity, $r+\alpha+t=1$, where $r$, $\alpha$ and $t$ are the reflection, absorption and transmission coefficients respectively (Kirchoff's law). It can be easily demonstrated that for a grey body in thermal equilibrium it must be $\alpha=\epsilon$ (assuming zero transmission, $t=0$). This can easily be demonstrated from \Fref{fig:Kirchoff}: a grey body inserted into a blackbody of a given temperature must eventually also reach at equilibrium the same temperature (2nd principle of thermodynamics), and in such a case the absorbed and emitted radiations must be equal. We can then conclude that $r=1-\alpha=1-\epsilon$. An object with poor reflectivity will have a high emissivity (a high absorption), possibly being also visually "black".

The interaction of electromagnetic waves with surfaces is described by the surface resistance, as discussed in \Sref{sec:Rs}. In the case of metals one could also prove the Hagen-Rubens law:
\begin{equation} 
	r=1-4\frac{R_s}{R_0}
	\label{eq:HagenRubens}
\end{equation}
where $R_s$ is the surface resistance of the metal surface and $R_0$ is the free space impedance (resistance) equal to 376.7~$\Omega$. From \Eref{eq:HagenRubens} we deduce that $\epsilon=4R_s/R_0$. At typical radio-frequencies the value of $\epsilon$ is extremely small (as can be deduced from the values discussed in the previous sections), thus the reflection coefficient of an RF electromagnetic wave from a metallic surface is very close to unity. At higher frequency the surface resistance should increase according to \Eref{eq:Rs}, but as mentioned in \Sref{sec:ASE} it stabilizes when attaining the anomalous skin effect regime. However at infrared frequencies and above the exact formulation is more complex. The considerations done in the preceding sections are no longer valid when $\omega \tau \gg 1$, which is typical at optical frequency (the reader could easily verify, using the value of $\tau$ mentioned earlier for copper at room temperature, that $\omega \tau = 1$ at a wavelength of incident radiation of 47~$\mu$m, in the far infrared). Rather surprisingly, in the regime $\omega \tau \gg 1$ the simple expression for the surface resistance \Eref{eq:Rs} becomes again valid. The emissivity of a metal in the infrared and visible spectrum, which are the most interesting for radiative heat exchanges, depends then only on its electrical resistivity. Good conductors such as copper have emissivity of the order of 0.03 (at 10~$\mu$m wavelength), while stainless steel can have an emissivity of around 0.2.

\begin{figure}
	\centering\includegraphics[width=.7\linewidth]{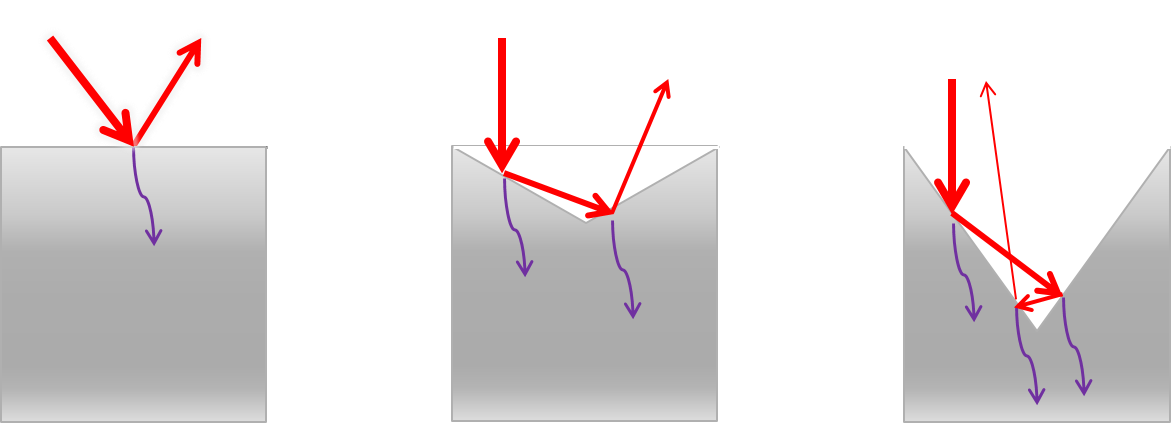}
	\caption{Pictorial illustration of the effect of roughness, where multiple reflections may reduce the total reflectivity of a surface.}
	\label{fig:Roughness}
\end{figure}

\begin{figure}
	\centering\includegraphics[width=.9\linewidth]{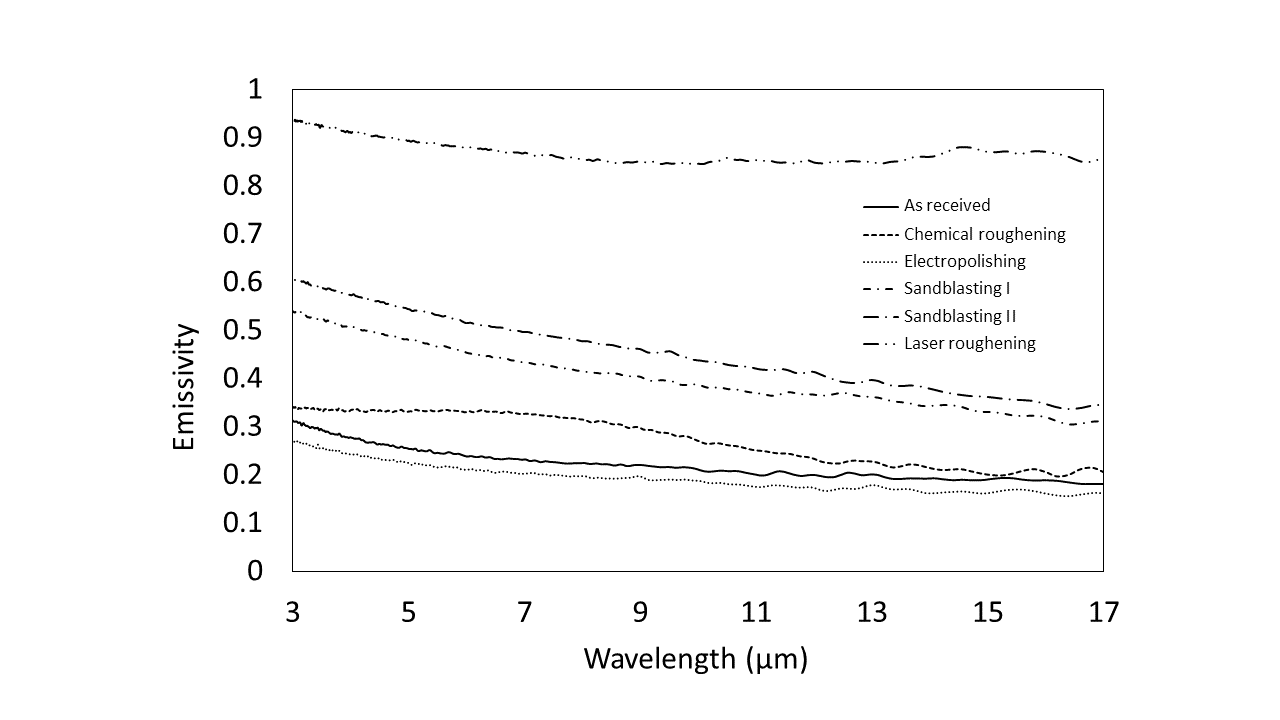}
	\caption{Effect of topographic roughness on the spectral emissivity of stainless steel: continuous line is as-received, dotted line is with a smoothening electropolishing treatment, dashed line is with a chemical roughening treatment, dotted-dashed lines are with different roughening sandblasting treatments, double-dotted-dashed line is with a laser roughening treatment.}
	\label{fig:Laser}
\end{figure}

One has to observe that the reflectivity depends also on the angle of incidence of the radiation onto the surface. The study of more complex cases than normal incidence goes beyond the scope of this lecture, we should nevertheless underline that in heat exchange processes the electromagnetic radiation is absorbed and reflected from any direction. Integrating the process over a full hemisphere results in the so-called hemispherical emissivity, which is the value often found in data tables. It is also clear from this that the surface state can play an important role, in particular its roughness. Emissivity values are often quoted for flat polished surfaces, a higher roughness resulting in a higher emissivity (less reflection). This is illustrated pictorially in \Fref{fig:Roughness}, which we leave intentionally uncommented, noting only that emissivity values close to unity can be obtained by means of suitable surface treatments, as illustrated in \Fref{fig:Laser}.

In practical terms, recalling the examples mentioned at the beginning of this section: in order to properly allow for cooling by radiation of objects which might get heated in vacuum, a high emissivity is needed; in order to properly thermally shield cryogenic components which might get heated from radiation coming from the external environment, a high reflectivity (low emissivity) is needed.

In some cases, different emissivity values are needed at different wavelengths. Although not related to particle accelerators, we can mention the case of solar panels which have a high emissivity in the visible spectrum in order to capture as much of the solar energy as possible, and a low emissivity in the infrared spectrum, in order to radiate away as little heat as possible. This is usually achieved by thin film coatings of high emissivity in the visible spectrum on components made of copper, which as we have seen has a low infrared emissivity.

\subsection{Contact resistance}
\label{sec:contacts}
\begin{figure}
	\centering\includegraphics[width=.6\linewidth]{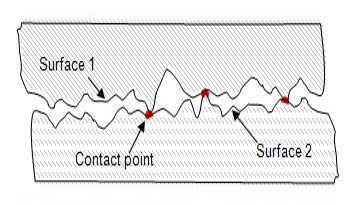}
	\caption{Pictorial illustration of how surface roughness may affect contact resistance.}
	\label{fig:Contact}
\end{figure}

Accelerators components in reality never "float" in vacuum (the particle beam does!), thus heat exchange by simple contact to other objects occurs naturally. However this seemingly trivial fact does not have a simple quantitative description. A hint for the explanation comes from \Fref{fig:Contact}. Real surfaces are in contact only in a very limited number of regions and the effective contact area $A$ is typically $A \propto p^n$ where $p$ is the contact pressure and the exponent $n$ is of the order of $0.5\div 1$, the thermal resistance in first approximation being inversely proportional to the contact area. The exponent depends on the mechanical properties of the materials, whether the parts under contact are in elastic or plastic regime, the "shape", height and pitch of the roughness, $\dots$. The proportionality in itself depends on whether the materials have or not an oxide, its thickness, the presence of possible contamination layers, the nature of the materials themselves, $\dots$. For all these reasons, it is not possible to give a general law, and all critical engineering designs relying upon thermal contacts in vacuum must be validated by experiments.  As an order of magnitude, typical values of thermal contact resistance between metals in vacuum can be of the order of $10^{-3}\ \text{m}^2$K/W at 1~bar of pressure, values that can be one order of magnitude larger or smaller depending on the materials and surface finish, and which vary of course with contact pressure as seen before. 

A reasoning similar to the one just made for thermal contacts can also be applied to electrical contacts, for which it is common practice to make contacts out "noble" metals which have no surface oxide, possibly using a "soft" material like gold and taking care of having smooth and clean surfaces. A good example are the RF fingers in the LHC (the flexible components allowing joining adjacent vacuum chambers and allowing for thermal expansion and geometrical adjustment of the machine), whose contacts are made of a fixed part which is rhodium-plated (for its smoothness and hardness) and another which is gold-plated (soft). These excellent contacts achieve a contact resistance of about $10^{-9}\ \Omega \text{m}^{-2}$, with a contact pressure of the order of 1~bar.

The detailed mechanism of heat transfer between the points of contact of two different materials goes beyond the scope of this lecture, but we can just mention that this is mostly due to electron and phonon tunnelling across the junction, both mechanisms having peculiar temperature dependencies as we will see in the next section.

\section{Thermal conductivity}
\begin{figure}
	\centering\includegraphics[width=.7\linewidth]{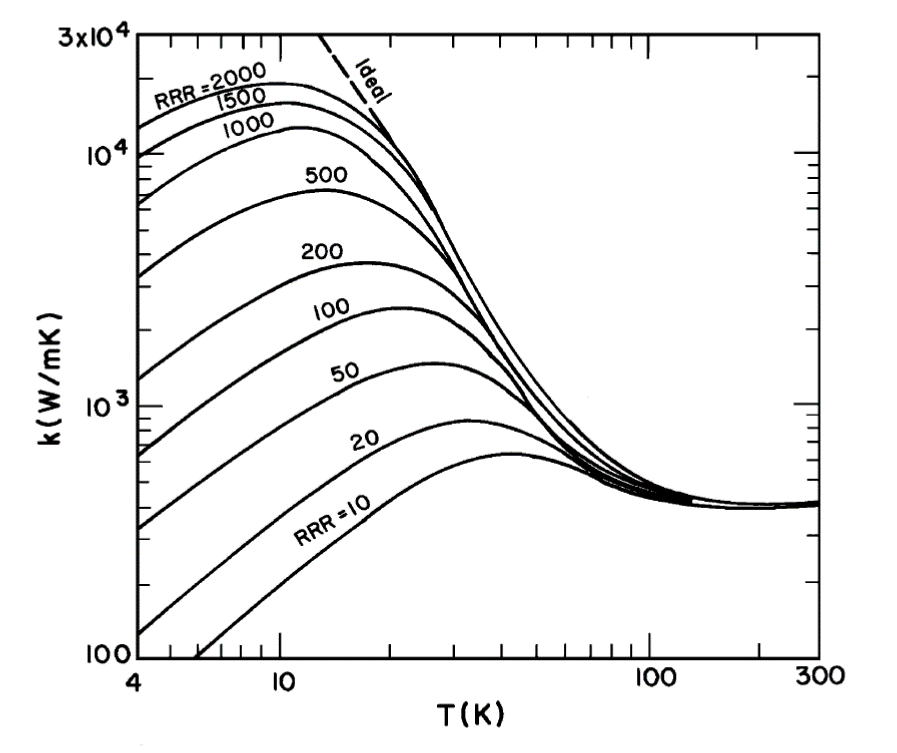}
	\caption{Thermal conductivity of copper as a function of temperature, for different values of RRR.}
	\label{fig:ThermCondCu}
\end{figure}

Thermal conductivity in a solid is one of the most complex phenomena to describe, and its study was actually a stepping stone in the development of the theory of phonons. Nevertheless, both phonons and electrons contribute to thermal conductivity in metals and their action together gives rise to its peculiar temperature dependence, as can be seen in \Fref{fig:ThermCondCu}, with a peak at low temperatures which depends on purity and a decrease as $T$ below it, while at higher temperature it stabilizes to a constant value. In order to understand the origin of this behaviour, we have first to discuss thermal conductivity in insulators.

\subsection{Thermal conductivity of insulators}
Heat conduction in insulators is a statistical process, and is due to phonons. To describe it, we will make an analogy with gases, where the kinetic theory predicts a link between thermal conductivity $K$ and specific heat $C$:
\begin{equation} 
	K=\frac{1}{3}C v \ell
	\label{eq:K_C}
\end{equation}
where $v$ is the speed of the particles and $\ell$ the mean free path between collisions. We recall that for an ideal gas the energy of the particles is equal to $k_B T$ per degree of freedom.

We have discussed in \Sref{sec:e-ph} that the number of phonons excited in a crystal lattice depends on temperature. For temperatures $T\gg \Theta_D$ it results also that $k_B T / \hbar$ is large compared to all phonon frequencies, i.e. every normal mode is in a highly excited state. In this case, the behaviour is quasi-classical and the total energy of the crystal is $3N\ k_B T$. The specific heat then reduces to $k_B$ times the density of normal modes, $C_{ph}=k_B\ 3N/V$.  This is  the classical Dulong-Petit law. At high temperatures the number of phonons is proportional to $T$. The mean free path of a phonon should be inversely proportional to the number of phonons with which it can interact, thus $\ell \propto 1/T$. From \Eref{eq:K_C} we can thus conclude that for $T\gg \Theta_D$ the thermal conductivity due to phonons is $K_{ph} \propto 1/T$.

For temperatures $T \ll \Theta_D$ only a small fraction of normal modes are excited, proportional to $(T/\Theta_D)^3$ (this is linked to the volume of the momentum-space of the phonons). The resulting specific heat is $\displaystyle \frac{12\pi^4}{5}\frac{N}{V}k_B\left( \frac{T}{\Theta_D}\right) ^3$. At low temperatures the number of phonons is small, thus their mean free path is limited only by collisions with impurities (or with grain boundaries, etc.), process which does not depend on temperature. From \Eref{eq:K_C} we can thus conclude that for $T \ll \Theta_D$ the thermal conductivity due to phonons $K_{ph}\propto T^3$.

\subsection{Thermal conductivity of metals}
The thermal conductivity of metals is the sum of the conductivities of phonons, like in an insulator, and of electrons.

The thermal conductivity $K_{el}$ of electrons can also be derived from \Eref{eq:K_C} where in this case the specific heat of the electron (Fermi) gas is $\displaystyle \frac{\pi^2 n k_B^2 T}{m v_F^2}$. We can thus conclude that:
\begin{equation} 
	K_{el}=\frac{1}{3}C_{el} v_F \ell = \frac{\pi^2 n k_B^2 T}{3 m v_F^2} v_F \ell = \frac{\pi^2 n k_B^2 T}{3 m} \tau .
	\label{eq:C_el}
\end{equation}
Recalling also \Eref{eq:B-G} we can see that for $T \gg \Theta_D$ we have $\tau \propto 1/T$ because of electron-phonon collisions, while for $T \ll \Theta_D$ we have $\tau=const$ because it is limited by defects only. Hence the electron thermal conductivity $K_{el}$ does not depend on $T$ for $T \gg \Theta_D$ , while it varies as $T$ for $T \ll \Theta_D$.

\begin{figure}
	\centering\includegraphics[width=.6\linewidth]{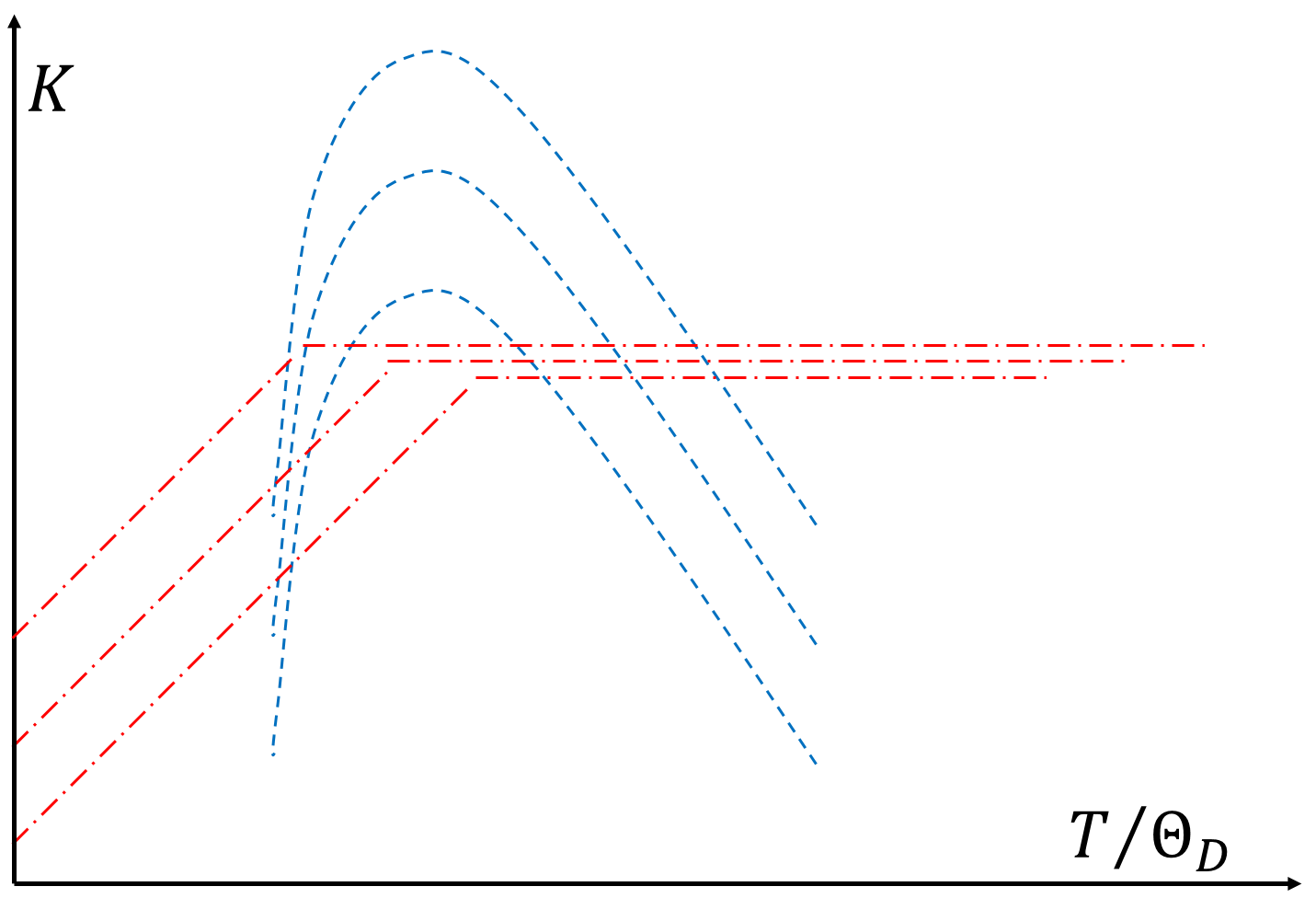}
	\caption{Pictorial representation of electron (dot-dashed) and phonon (dashed) thermal conductivities of copper (compare with \Fref{fig:ThermCondCu}), for different values of RRR.}
	\label{fig:ThermCond}
\end{figure}

All the above information is summarised in \Fref{fig:ThermCond}, which can be compared to \Fref{fig:ThermCondCu} for copper. The electron thermal conductivity as well as the phonon thermal conductivity have a common dependence on purity at low temperature, while at high temperature where only electron thermal conductivity plays a role, the thermal conductivity converges to a unique and constant value.

\subsubsection{The Wiedemann-Franz law}
\label{WiedemannFranz}
It is intuitive, when comparing \Fref{fig:Cu_RRR} and \Fref{fig:ThermCondCu} to conclude that there is a proportionality relationship between electrical and thermal conductivities for $T \gg \Theta_D$, i.e. when both are dominated by electron-phonon collisions. This is the so-called Wiedemann-Franz law, and it is expressed as:
\begin{equation} 
	\frac{K_{el}}{\sigma} = L T
	\label{eq:W-F}
\end{equation}
where $L$ is called Lorenz number and it is equal to $2.45\times10^{-8} \ \text{W} \Omega \text{K}^{-2}$ . Beyond eminently theoretical considerations, this equation is in fact extremely useful for carrying out simple estimations if only one quantity is know while the other is needed. This is true also in the case of alloys at high temperature, for which data sometimes are not readily available. The Wiedemann-Franz law can also be extended to electric contacts, as one could verify comparing the values of thermal and electric contact resistances in \Sref{sec:contacts}, although in this case the results must be considered only as rough approximations at best.

\section{Conclusions}
In this lecture we have condensed several different topics related to electrical and thermal conductivities of metals, seen from the perspective of their applications to particle accelerators, and in particular the fabrication of vacuum chambers and beam-facing components in vacuum.
This lecture cannot replace of course basic textbooks on solid-state physics and material science, to which the reader is referred for further information; nor the simple experimental values mentioned for a few selected cases can replace proper data sheets corresponding to the materials that one wishes to employ for any particular project.

\end{document}